\documentstyle[prl,aps,twocolumn,psfig]{revtex}

\begin{document}

\twocolumn[\hsize\textwidth\columnwidth\hsize\csname
@twocolumnfalse\endcsname

\draft

\title{Shearing a fermionic gas and quantized friction}
\author{Er'el Granot\footnotemark}

\address{Department of Electrical and Electronics Engineering, College of Judea and Samaria, Ariel 44837, Israel}

\maketitle
\begin{abstract}
\begin{quote}
\parbox{16 cm}{\small

The frictional forces in sheared fermionic gas are investigated.
The gas is sheared by two sliding surfaces. Except for a small
imperfection (bulge) on one of them, the surfaces are totally
smooth. We show that when the bulge is extremely small, and the
gas is also confined in the lateral dimension, the frictional
force ($F$) is \emph{quantized}. That is, $F=m\mu_Fv$, where $v$
is the sliding velocity with respect to the lubricant gas, $m$ is
an integer, and $\mu_F$ is the friction coefficient. It is also
shown that $\mu_F$ is proportional to the Planck constant $h$ and
depends only on the bulge's properties (it does not depend on
either the gas properties or the sliding velocity).
 }
\end{quote}
\end{abstract}


]

\narrowtext \footnotetext{erel@yosh.ac.il} \noindent

\section{introduction}

Recent developments in micro-electromechanical systems have raised
practical interest in research involving quantum friction
\cite{Rubio_96,Kardar_99,Maclay_01,Chen_Mohideen_02,Chen_Mohideen_02B,Serry_W_1995,Chan_01,1_Granick_99,2_Fuhrmann_98,3_Burns_99,4_Liebsch_97}.
However, the investigation of quantum friction has revealed some
surprising fundamental effects. For example, two smooth dielectric
surfaces moving laterally (parallel to one another) experience
frictional forces as if the vacuum between them were a viscous
fluid\cite{6_Pendry_97,5_Pendry_98,Volokitin_Presson_99}.
Therefore, friction measurements can contain a great deal of
information on the nature of vacuum fluctuations. Sliding
surfaces, however, are seldom smooth; in fact, they are usually
rough and corrugated. The most common remedy for such roughness
is, of course, lubrication, and in modern micro-machinery, the
lubricants are usually gases\cite{Li_05,Chen_04}. The possibility
of confining a fermionic gas in space with a laser beam has been
shown experimentally (see, for example,
refs.\cite{Ohara_etal_99,7_Zaloj_99}).

In this paper, we investigate the onset of frictional forces
between two adjacent surfaces. The two surfaces are sliding
against one another at zero temperature, whilst the shearing is
lubricated by a fermion (say, atoms or electrons) gas. On one of
the surfaces there is a small bulge. The collisions between the
particles and the bulge create the frictional force. The main
argument of (and motivation for) this paper is the following: a
small fraction from each of the transversal modes is reflected
from the bulge. The modes, which are propagating against the
bulge, are reflected from it with a slightly higher energy. When
the sliding velocity is low (in comparison to the Fermi velocity),
the energy increase is proportional to the particles' momentum,
and therefore, the power activated upon the bulge is proportional
to $p^2R$ ($p$ is the momentum and $R$ is the modes' reflectivity
coefficient-- see a detailed discussion further on). Now, since
for a very weak protrusion $R\sim p^{-2}$, one can conjecture that
each mode makes \emph{exactly} the same contribution to the
frictional force, which therefore should be quantized. In this
paper, we show that for a very weak protrusion (the bulge), this
is indeed the case.

\begin{figure}
\psfig{figure=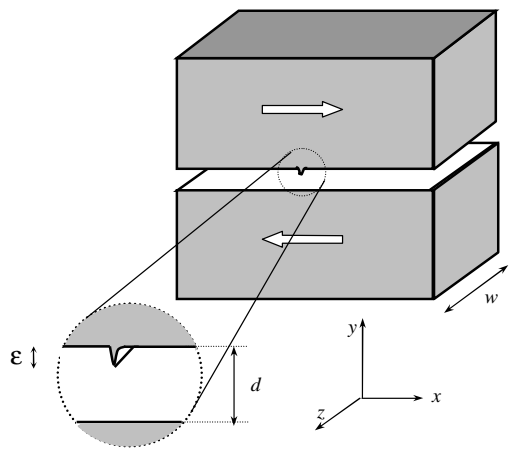,width=10cm,bbllx=100bp,bblly=370bp,bburx=550bp,bbury=660bp,clip=}
\caption{\emph{An illustration of the two sliding surfaces. On the
upper surface there is a small bulge, which is magnified in the
inset. The fermionic gas is confined between the two surfaces.
}}\label{fig1}
\end{figure}

\section{The system's description}

Imagine a lubricating fermionic gas confined between two
infinitely long surfaces at zero temperature (they are infinitely
long in the $x$-direction and have a finite width in the $z$
direction, see Fig.1). The fermionic gas is thus confined to a
rectangular cross section, whose width corresponds to the
surfaces' width ($w$) and whose height is equal to the distance
between them ($d$ in the $y$-direction). In the following, we
discuss only very close surfaces, i.e., $d\ll w$. On one of the
surfaces, say the upper one, there is a small protrusion (bulge or
bump, for example, see Fig.1). For simplicity, it is assumed that
the protrusion has no features in the transverse direction, i.e.,
it is independent of $z$, and thus can be fully characterized by
its cross section in the $x$-$y$ plane. When the two surfaces
slide against each other, with a relative velocity $2v$ (i.e.,
each of the surfaces slides with a velocity $v$ with respect to
the lubricant gas), the protrusion's resistance to the sliding is
manifested by a frictional force. If not for the bulge, the
shearing process would not influence the lubricant. Classically,
the particles of the lubricant gas, which are bouncing bouncing
back and forth in their confinement zone prevent the bulge from
sliding smoothly through the lubricant.

The Schr\"{o}dinger equation, which describes the dynamics of the
confined lubricant, reads
\begin{equation}
-\frac{\hbar^2}{2m_0}\nabla^2\psi+
\left[V_{su}(y)+V_{co}(z)+V_{pr}(x,y)\right]\psi=i\hbar\frac{\partial\psi}{\partial
t} \label{shro_eq}
\end{equation}

where $V_{su}(y)$ is the potential of the surfaces,
$V_{co}(z)\equiv \left\{
\begin{array}{cc}
  0 & 0\leq z\leq w \\
  \infty & otherwise
\end{array} \right.$ is the confinement potential in the $z$-direction and
$V_{pr}(x,y)$ is the potential of the protrusion, which satisfies
$V_{pr} \rightarrow \infty$ for $|x|\rightarrow \infty$; note that
it does not depend on $z$.

In the following discussion, we assume zero temperature, and a
very sparse lubricant (very low density), i.e., the density per
unit length ($n$) satisfies $n\ll d^{-1}$ . Therefore, only the
first transversal mode (which is related to the coordinate $y$) is
occupied. In the $z$-direction, however, many modes can survive,
which will be identified by their quantum number $j$. For any
given energy $E=\hbar\omega$ and a given channel $j$, the quantum
wave function of the lubricant particles can be written:

\begin{equation}
\psi^{\pm}_{p,\omega}=\sin(\pi y/d)\sin(\pi z p/w)\exp(\pm
ik^jx-i\omega t) \label{modes}
\end{equation}
where the longitudinal wave number $k^j$ satisfies: $0\leq k^j
\leq k^j_F$ , when

\begin{equation}
\left(k^j_F \right)^2\equiv 2m_0E_F/\hbar^2-(\pi/d)^2-(\pi j/w)^2
\label{K_fermi}
\end{equation}

$m_0$ is the lubricant gas particles' mass; $j$ is an integer that
characterizes the transversal mode, $E_F$ is the Fermi energy and
the upper (lower) sign in eq.(\ref{modes}) refers to particles
which propagate from the left (right) side of the bulge to its
right (left) side.

In order to solve this problem, it is very convenient to choose a
frame of reference in which the bulge is at rest. Consequently,
the bulge "sees" (in the new frame of reference) on its right,
particles emerging with the wave numbers $0 \leq k_r \leq
k^j_F+m_0v/\hbar$, for every channel $j$, while on its left the
incoming particles have, for the same channel, the wave numbers:
$0 \leq k_l \leq k^j_F-m_0v/\hbar$ (see Fig.2).

In the following we discuss only very low sliding velocities,
i.e., $mv/\hbar \ll k^j_F$ , which means that the particles'
maximum energy on the right is higher by $\Delta\varepsilon_j
\cong 2k^j_F v\hbar$ (in the moving frame) than the particles'
maximum energy on the left.

\begin{figure}
\psfig{figure=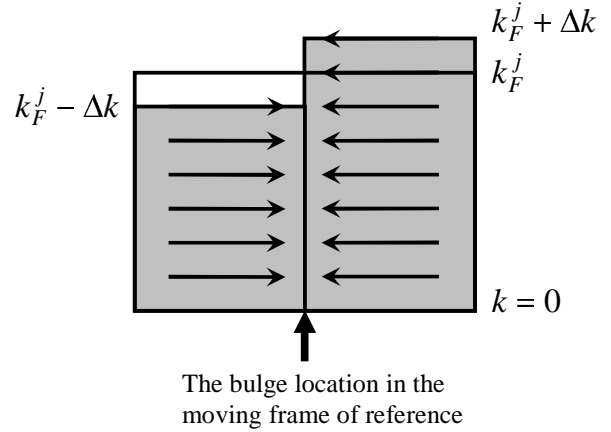,width=10cm,bbllx=90bp,bblly=460bp,bburx=490bp,bbury=740bp,clip=}
\caption{\emph{At the moving frame (where the bulge is at rest),
the particles on both sides of the bulge can be regarded as
hitting it with slightly different momenta (for the $j$th mode in
the figure), $\Delta k \equiv m_0v/\hbar$.}}\label{fig2}
\end{figure}

Clearly, in the first approximation, only the particles that
propagates from the right (to the left) with momentum
$k^j_F-m_0v/\hbar\leq k_r \leq k^j_F+m_0v/\hbar$ contribute to the
energy wasted by friction.

Let us denote by $T_j\equiv T(k=k^j_F)$ the quantum transmission
(for transport in the $x$-direction) through the bulge via the
$j$th channel. It should be pointed out here that it is possible
to describe the transmission by a single parameter since the
geometry of the bulge cannot allow for transmission from one
channel to another. The reflected current of the $j$th channel
(note that here we are regarding a particle's current and not a
charge current) from the bulge, which includes only the particles
with $k_r \geq k^j_F-mv/\hbar$, is equal to

\begin{equation}
I_j=\Delta \varepsilon_j(1-T_j)/\hbar \label{'I_j'}
\end{equation}.

Each of the reflected particles gains an energy quantum
$2\Delta\varepsilon_j$, and therefore the power imposed upon the
bulge is

\begin{equation}
P=2\sum_j \Delta\varepsilon_j^2(1-T_j)/\hbar \label{power}
\end{equation}.

Therefore, the power produced during the shearing process
maintains the relation

\begin{equation}
P=\frac{4}{\pi}v^2\hbar \sum_j (k^j_F)^2(1-T_j) \label{power2}
\end{equation}.

Finally, the friction force reads

\begin{equation}
F=dP/dv=\mu_F v \label{fric_force}
\end{equation}.

where the friction coefficient is
\begin{equation}
\mu_F \equiv h\sum_j n_j^2(1-T_j) \label{fric_coef}
\end{equation}

and $n_j\equiv (2/\pi)k_F^j$ is the particles' density per unit
length of the $j$th channel (the prefix 2 comes from the spin
degeneracy).

Eq. \ref{fric_force} suggests a linear relationship between the
frictional force and the shearing velocity. Notice that $T_j$ is
also a function of the lubricant density, since it is a function
of the Fermi wave number. On the other hand, the transmission
coefficient $T$ has no dependence on the lubricant's
characteristics other than its density, but it is highly sensitive
to the bulge's properties, and therefore no general relation can
be obtained. However, a general conclusion that does arise from
Eq. \ref{fric_coef} is that

\begin{equation}
\mu_F \leq \mu_F^{max} \equiv h\sum_j n_j^2 \label{mu_max}
\end{equation}

regardless of the bulge's characteristics or dimensions. Moreover,
this result even holds true for an arbitrary number of
protrusions, and thus can be applied to any rough surface. For any
number of channels, Eq. \ref{mu_max} can be evaluated as

\begin{equation}
\mu_F^{max}=h\left[\left(q^2-\frac{4}{d^2}
\right)m-\frac{2}{w^2}\frac{m(m+1)(2m+1)}{3} \right]
\label{mu_max2} \end{equation}

where $q^2\equiv (2m_0E_F/\hbar^2)(4/\pi^2)$ and $m$, which is the
integral part of $w\left(q^2/4-1/d^2 \right)^{1/2}$, is the
channel number (note the difference between the channel number $m$
and the lubricant particle mass $m_0$). When $m \rightarrow
\infty$, i.e., $w \rightarrow \infty$, it can be approximated by

\begin{equation}
\frac{\mu_F^{max}}{w}=\frac{1}{3}h\left(q^2-4/d^2\right)^{3/2}
\label{mumax_infty}
\end{equation}

When the onset of friction is being considered the exact nature of
the protrusion is very important. Of course, the transmission $T$
cannot have a general expression; however, in a wide range of thin
and small bulges (or bumps), the transmission can be expressed by
\cite{9_Granot_99}

\begin{equation}
T_j \cong \left(1+(n_c/n_j)^2 \right)^{-1} \label{equation}
\end{equation}

where $n_c$ is a transition density, which characterizes the
bulge. If the problem was a one-dimensional one, and the
protrusion was a delta function, then the approximation sign
'$\cong$' would be replaced by the equality sign '$=$ '.
Nevertheless, if the protrusion is a very small bulge, this
expression can be applied with great accuracy, even to problems
with more than a single dimension (see refs. \cite{9_Granot_99}).

\begin{figure}
\psfig{figure=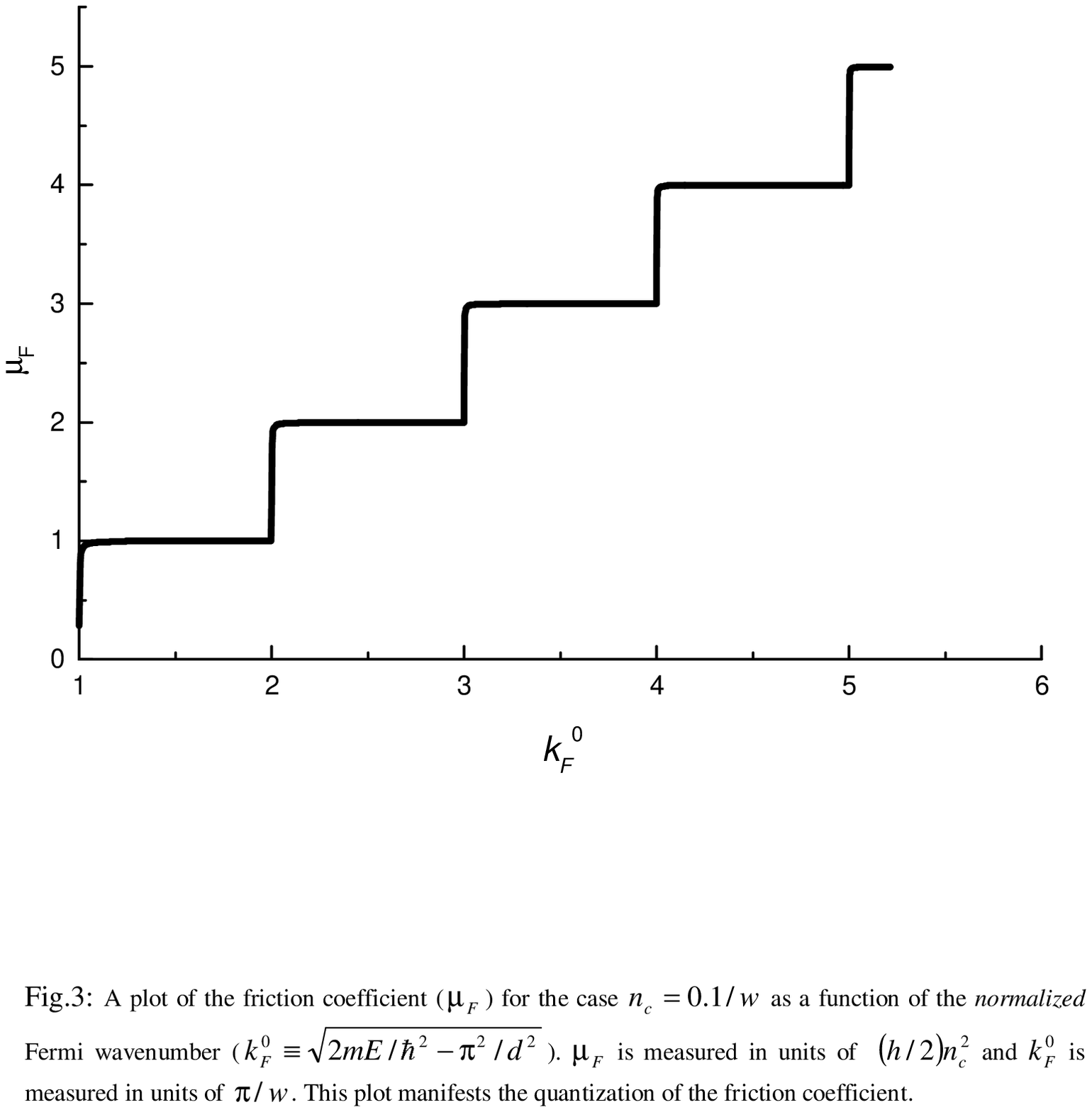,width=10cm,bbllx=60bp,bblly=330bp,bburx=600bp,bbury=690bp,clip=}
\caption{\emph{A plot of the friction coefficient ($\mu_F$) for
the case $n_c=0.1/w$ as a function of the normalized Fermi wave
number ($k_F^0\equiv \sqrt{2m_0E/\hbar^2-\pi^2/d^2}$). $\mu_F$ is
measured in units of $hn_c^2$ and $k_F^0$ is measured in units of
$\pi/w$. This plot manifests the quantization of the friction
coefficient.}}\label{fig3}
\end{figure}

If, for example, the protrusion is a very shallow but long bulge,
then $n_c=L^{-1}$, where $L$ is the bulge's length. If the bulge
is a point impurity, then $n_c=\nu \varepsilon^2/d^3$, where
$\varepsilon$ is the distance from the surface and $\nu$ is a
parameter that depends on the eigen-energy of the impurity, and
has a weak dependence on $\varepsilon$ (see refs.
\cite{9_Granot_99,9_Granot_04_05}).

In any case, Eq. \ref{fric_coef} should read
\begin{equation}
\mu_F=hn_c^2\sum_j{\frac{1}{1+(n_c/n_j)^2}} \label{fric_coef2}
\end{equation}

For very sparse lubricant ($n_j \ll n_c$ for every channel $j$),
the friction coefficient expectedly vanishes and the expression in
eq. \ref{mu_max} is regained: $\mu_F=h\sum_j n_j^2$.

More important is the case of a point protrusion, that is
$\varepsilon \rightarrow 0$, or $n_c \ll n$. This regime is
important not only because it investigates \emph{the onset of
friction} due to a minuscule defect (bulge), but also because it
even applies to relatively high densities (so long as the distance
between the surfaces is small enough, i.e., $n_j \ll d^{-1}$ for
every $j$), in which case the interaction between the particles
(if there is any) can be neglected.

The friction coefficient (eq. \ref{fric_coef2}) for the case
$n_c=0.1/w$ as a function of the Fermi wave number $k_F^0 \equiv
\sqrt{2m_0E_F/\hbar^2-\pi^2/d^2}$ (see definition in Eq.
\ref{modes}) is presented in Fig.3. $\mu_F$ is measured in units
of $hn_c^2$ and $k_F^0$ is measured in units of $\pi/w$. The plot
exhibits a staircase pattern and reveals the \emph{quantization}
of the friction coefficient $\mu_F$, where the friction
coefficient quantum is $hn_c^2$. This quantity does not depend on
either the lubricant's properties or its density. It depends only
on properties of the bulge and the distance between the surfaces.

When $n_c \rightarrow 0$, the staircase pattern is more sharply
delineated, and Eq. \ref{fric_coef2} can be written
\begin{equation}
\mu_F=mn_c^2h \label{fric_quantization}
\end{equation}

where $m$ is the total number of propagating channels. This is the
main result of this paper, which manifests the quantization of
quantum friction for extremely weak protrusions. Each mode has
exactly the same contribution to the friction force:
\begin{equation}
\delta F=n_c^2hv
\end{equation}

\section{Numerical Evaluation and Possible Implementation}

If the bulge is a point protrusion, whose de Broglie wavelength is
approximately equal to $\varepsilon$ and, for example, $d \sim
1nm$, $\varepsilon_0/d \sim 0.3$ and $v=1m/sec$, then the friction
quanta $\delta F$, (i.e., the force is $F=m\delta F$) is around
$\delta F>10^{-13}$N -- a small but measurable quantity
\cite{Chen_Mohideen_02,Chen_Mohideen_02B}.

Note that the friction quantum is independent (in the high-density
regime) of the gas properties. Therefore, when the density
increases, the force increases with it but not its quanta.

To enlarge $\delta F$, one can use more bulges or more parallel
channels (see Fig.4). Since the gas is confined only to the laser
beam, it is possible to use many parallel beams, each of which
makes the same contribution to the force (note that $\delta F$ is
independent of $w$).

\begin{figure}
\psfig{figure=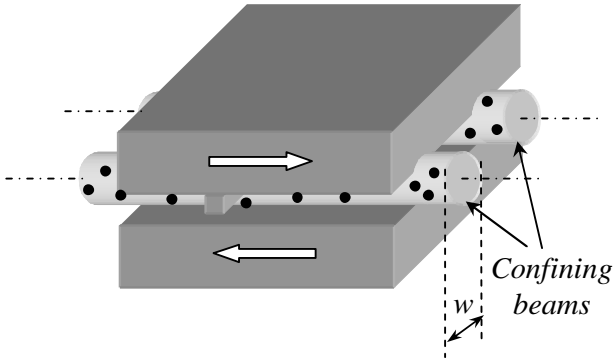,width=10cm,bbllx=123bp,bblly=468bp,bburx=420bp,bbury=660bp,clip=}
\caption{\emph{The gas is confined by laser beams. Each beam makes
the same contribution to the friction force. The black dots
represent the confined atoms (fermions). }}\label{fig4}
\end{figure}

\section{Summary}
The frictional force which emerges in a lubricant fermionic gas
due to shearing was investigated. The discussion focused on the
onset of the frictional force. It was shown that when the
smoothness of one of the sliding surfaces is damaged by a single
small imperfection, the frictional force is proportional to the
sliding velocity, i.e., $F=\mu_F v$, where the friction
coefficient is quantized $\mu_F=mn_c^2h$ ($m$ is an integer). The
friction coefficient quantum $\delta\mu_F=n_c^2h$ depends only on
the geometry, i.e., on the bulge and on the distance between the
surfaces; it does no depend on either the lubricant gas properties
or the sliding velocity.

I would like to thank prof. Mark Azbel for enlightening
discussions.

\end{document}